\documentclass[apj]{emulateapj}
\usepackage[backref,breaklinks,colorlinks,citecolor=blue]{hyperref}

\usepackage{amssymb}
\usepackage{amsmath}
\usepackage{amsthm}
\usepackage{nicefrac}
\usepackage{graphicx}
\usepackage{epstopdf}

\makeatletter
\newcommand*{\citenumns}[2][]{%
  \begingroup
  \let\NAT@mbox=\mbox
  \let\@cite\NAT@citenum
  \let\NAT@space\NAT@spacechar
  \let\NAT@super@kern\relax
  \renewcommand\NAT@open{}%
  \renewcommand\NAT@close{}%
  \cite[#1]{#2}%
  \endgroup
}
\makeatother

\begin{document}

\title{A lower bound on adiabatic heating of compressed turbulence \\ for simulation and model validation}

\author{Seth Davidovits\altaffilmark{$\dagger$}}
\author{Nathaniel J. Fisch}
\affiliation{Department of Astrophysical Sciences, Princeton University, Princeton, NJ 08544, USA }

\altaffiltext{$\dagger$}{Corresponding author: \href{mailto:sdavidov@princeton.edu}{sdavidov@princeton.edu}}

\begin{abstract}
The energy in turbulent flow can be amplified by compression, when the compression occurs on a timescale shorter than the turbulent dissipation time. This mechanism may play a part in sustaining turbulence in various astrophysical systems, including molecular clouds. The amount of turbulent amplification depends on the net effect of the compressive forcing and turbulent dissipation. By giving an argument for a bound on this dissipation, we give a lower bound for the scaling of the turbulent velocity with the compression ratio in compressed turbulence. That is, turbulence undergoing compression will be enhanced at least as much as the bound given here, subject to a set of caveats that will be outlined. Used as a validation check, this lower bound suggests that some models of compressing astrophysical turbulence are too dissipative. The technique used highlights the relationship between compressed turbulence and decaying turbulence.
\end{abstract}

\section{Introduction}
Turbulence undergoing mean compression, also called compressed turbulence, is of interest in a variety of disciplines. A number of studies, ranging from investigations of its essential behavior to detailed application studies, have been conducted with an eye towards internal combustion engines and aerodynamic flows. These include studies focusing on the zero-mach-number limit (e.g. ~\citet{morel1982,wu1985,coleman1991,cambon1992,guntsch1996,liu2010,hamlington2014}), and those focusing on the finite-mach-number limit (e.g. ~\citet{blaisdell1991,speziale1991,durbin1992,coleman1993,cambon1993,blaisdell1996,grigoriev2016}).

Other contexts where compressed turbulence is of interest include plasma physics and inertial fusion~\citep{davidovits2016a,davidovits2016b,thomas2012,weber2014,kroupp2011,kroupp2007,maron2013,kroupp2007a}, and astrophysics~\citep{robertson2012}. In the astrophysics context, the turbulence undergoing compression is typically supersonic, and the present work focuses on compressed turbulence in this context.

Turbulence is ubiquitous in interstellar gas~\citep{elmegreen2004}, and the properties of supersonic turbulence have been related to important astrophysical questions such as the core mass and stellar initial mass functions~\citep{padoan2002,ballesteros2006,hennenbelle2008}, star formation efficiency~\citep{elmegreen2008}, and the origin of Larson's laws~\citep{kritsuk2013a}. As such, supersonic turbulence has been the subject of numerous investigations in the context of astrophysics~(e.g. \citet{maclow1998,maclow1999,kritsuk2007,federrath2008,kritsuk2013b,federrath2013,banerjee2014}). This astrophysical turbulence is often undergoing contraction or expansion under the influence of gravity or pressure. \citet{robertson2012} pointed out that little work has been done on compressed turbulence in astrophysics, although intuition and some results from prior work on compressed turbulence in other contexts should be expected to carry over. Since contraction (or expansion) influences the behavior of the turbulence, and the turbulence plays a role in many problems related to interstellar gas dynamics, it is desirable to better understand how exactly contraction influences turbulent behavior.

At the most basic level, the first classifying parameter for turbulence undergoing compression is the ratio, $S = \tau_d/\tau_c$, of the turbulent dissipation time, $\tau_d$, to the compression time, $\tau_c$. When the compression is very slow, $S \ll 1$, and the compression has little effect. If the compression is very fast, $S \gg 1$, the turbulence is essentially ``frozen'' and its behavior can be treated with rapid distortion theory (RDT)~\citep{savill1987,hunt1990,durbin2010}. For three-dimensional rapid isotropic compressions, one finds that the root mean square (rms) turbulent velocity $v_{rms} \sim v_{rms,0}/\bar{L}$, where $\bar{L}$ is the contraction factor along each axis, $\bar{L} = L/L_0$ (see e.g. \citet{wu1985} for a zero-mach RDT treatment, or \citet{cambon1993} for a finite-mach RDT treatment; a similar result is given by~\citet{peebles1980} in section 90). In actuality, the turbulence will not be completely ``frozen'', and there will be turbulent dissipation, the quantity of which depends (in part) on the rapidity of the contraction. This dissipation reduces the rms turbulent velocity scaling with compression below the $1/\bar{L}$ ``adiabatic'' result.

Here we present an argument for an upper bound on the amount of this turbulent dissipation, thereby providing a lower bound on the amount of adiabatic heating that turbulence in a contracting gas can undergo. This argument rests on the following assumption. Consider as a base case the rate of decay for unforced Navier-Stokes (NS) turbulence with a constant viscosity. We assume that when the viscosity is a shrinking function of time, with the same initial value as the base case, the rate of decay is not larger than that for the base case. If this physically reasonable assumption holds, the bound follows directly. Then the bound can be used as a check on models and simulations of compressing turbulence, or as a model itself. We carry out an initial comparison with some previous work, which suggests that at least some approaches to simulating or modeling compressed high-mach turbulence are too dissipative. They will give, for example, asymptotic scaling (in $\bar{L} \rightarrow 0$) of the turbulent velocity for a gravitational contraction that is below the minimum predicted by the bound. Since similar approaches are used in many astrophysical simulations, this apparent disagreement with the bound may have implications for other work as well.

The focus of the current work is to present the bound and an initial comparison against some previous work, thereby motivating future work to determine if the key assumption made in arriving at the bound holds. We note that even if some approaches to simulating and/or modeling compressed high-mach turbulence are in fact too dissipative, a separate determination needs to be made as to whether this affects the results of interest. While physically reasonable, the assumption is not rigorous. In the surprising event the assumption is violated, so that the decay rate of NS turbulence is increased if the viscosity shrinks in time, there will likely still be implications for turbulence in astrophysical settings. Of course, the assumption (and bound) may hold in some situations and not in others, depending on the mechanism(s) by which the assumption is violated, if it is.

In the process of arriving at the bound, the sometimes forgotten relationship~\citep{cambon1992} between turbulence forced by contraction and decaying turbulence is highlighted. Beyond its use in the argument for a lower bound, which is the focus of the present work, this connection may be helpful for understanding the influence of contraction on astrophysical processes, since it gives a means of translating quantities (e.g. correlation functions) between compressing and decaying cases. The relationship can also be useful for simplifying simulations of compressing turbulence (e.g. as used by~\citet{davidovits2016a}).

Although the bound presented here has a number of caveats associated with it, the approach used to arrive at it should be adaptable to create new bounds with different applicability. The bound is given in terms of $\bar{L}$, the Hubble parameter, $H = \dot{L}/L$ (with the overdot the time derivative), and the decay time constant $t_0$ and power $\alpha$ (in the spirit of ~\citet{maclow1998,maclow1999}) for the rms velocity in decaying supersonic turbulence. The bound is
\begin{equation}
\frac{v_{rms}}{v_{rms,0}} \geq \frac{1}{\bar{L}} \left(1 + \frac{1}{t_0} \int_1^{\bar{L}} (\bar{L}')^{-3} \frac{\mathrm{d}\bar{L'}}{H'} \right)^{-\alpha/2}.\label{eq:bound}
\end{equation}
As will be shown later, this form of the bound follows once a fit for the decay of $v_{rms}$ in unforced NS turbulence (with a regular, constant viscosity) is chosen. If these fits are refined, the bound will be as well.

The paper is organized as follows. The model, essentially the NS equations in coordinates comoving with the contraction (or expansion), is described in Sec.~\ref{sec:model}. Section \ref{sec:rescaling} shows the use of a time-dependent variable rescaling to change the NS equations forced by contraction into NS equations for decaying turbulence, with extra time-dependent coefficients. An argument for the bound, Eq.~(\ref{eq:bound}), is given in Sec.~\ref{sec:bound}, using the rescaled NS equations. Section \ref{sec:discussion} compares the bound to some previous results on compressing supersonic turbulence and discusses the caveats and implications of the bound and rescaling.

\section{Model} \label{sec:model}
The model is the NS equations written in contracting (or expanding) coordinates. These coordinates, $\mathbf{x}$, are defined in terms of the proper coordinates, $\mathbf{r}$, as
\begin{equation}
\mathbf{x} = \mathbf{r}/\bar{L}. \label{eq:coord}
\end{equation}

The proper velocity, $\mathbf{u}$, written in terms of the peculiar velocity $\mathbf{v}$ and the contracting coordinates, is
\begin{equation}
\mathbf{u} = \dot{\bar{L}} \mathbf{x} + \mathbf{v} \! \left( \mathbf{x}, t \right) \label{eq:u}
\end{equation}
Beginning with the NS equations for $\mathbf{u}$ and the density $\rho$ in the proper coordinates, and rewriting in terms of $\mathbf{x}$ and $\mathbf{v}$, gives
\begin{eqnarray}
\frac{\partial \rho}{\partial t} + \frac{1}{\bar{L}} \nabla \cdot \left( \rho \mathbf{v} \right) + 3 \frac{\dot{L}}{L} \rho & = & 0 \label{eq:density}
\\
\frac{\partial \mathbf{v}}{\partial t} + \frac{1}{\bar{L}} \! \left( \mathbf{v} \cdot \nabla \right) \! \mathbf{v} + \frac{1}{\rho \bar{L}} \nabla p + \frac{1}{\bar{L}} \nabla \Phi + \ddot{\bar{L}} \mathbf{x} +\frac{\dot{L}}{L} \mathbf{v} & = & \frac{\mathbf{D}}{\bar{L}^2} \label{eq:momentum}
\\ 
\frac{1}{\rho} \left(\mu \nabla^2 \mathbf{v} + \left(\mu + \lambda \right)\nabla \left(\nabla \cdot \mathbf{v} \right) \right) & = & \mathbf{D}. \label{eq:dissipation}
\end{eqnarray}
Here, $p$ is the pressure, $\Phi$ is the gravitational potential, and $\mathbf{D}$ is the usual dissipation term in the momentum equation, which is given in Eq.~(\ref{eq:dissipation}). It has been assumed that the dynamic and bulk viscosities, $\mu$ and $\lambda$ respectively, are constants. A derivation of these equations, with the exception of the dissipation term, can be found in \citet{peebles1980}, section 9. Essentially identical equations, based on contractions identical to those dictated by $\mathbf{u}$ in Eq.~(\ref{eq:u}), but without the gravitational potential, underlie studies of compressing turbulence in other contexts \citep{blaisdell1991,cambon1993,coleman1993}.

Besides giving spatial derivatives time-dependent coefficients (powers of $\bar{L}$), the effect of the contraction is to add forcing (or dissipation) to both the density and momentum equations. In the density equation, Eq.~(\ref{eq:density}), the third term is a forcing term when $\dot{L}$ is negative. This, in part, causes the mean density to increase as expected for the contraction. 

In the momentum equation, Eq.~(\ref{eq:momentum}), the first term to the left of the equals sign is similarly a forcing term when $\dot{L}$ is negative. In fact, a similar term has been used as a way to add real space forcing for turbulence simulations~\citep{lundgren2003,rosales2005,petersen2010}. It is this term that, taken alone, will lead to the ``adiabatic'' increase of turbulent velocity $v_{rms} \sim 1/\bar{L}$. 

The second term to the left of the equality in Eq.~(\ref{eq:momentum}) is related to the acceleration of the contraction, $\ddot{\bar{L}}$. It depends on $\mathbf{x}$, and can cause the turbulence to be inhomogeneous (see \citet{blaisdell1991}, Section 2.4, for a thorough discussion). In the case where the contraction (the time dependence of $L$) is determined by gravity, this acceleration term can be removed from the momentum equation by the gravitational field of the mean density (see \citet{peebles1980}). 

For the present work, we will treat this as the case, and we will also choose to ignore the gravitational effects associated with density fluctuations (as in \cite{robertson2012}). The pressure is taken to obey a polytropic law,
\begin{equation}
p = K \rho^{\gamma}, \label{eq:polytropic}
\end{equation}
with $K$ and $\gamma$ constants. Together, these choices give the model momentum equation,
\begin{equation}
\frac{\partial \mathbf{v}}{\partial t} + \frac{1}{\bar{L}} \! \left( \mathbf{v} \cdot \nabla \right) \! \mathbf{v} + \frac{K}{\rho \bar{L}} \nabla \rho^{\gamma} + \frac{\dot{L}}{L} \mathbf{v} = \frac{\mathbf{D}}{\bar{L}^2}. \label{eq:f_momentum}
\end{equation}

\section{Rescaling} \label{sec:rescaling}

Substituting rescaled values of the density, velocity, and time,
\begin{eqnarray}
\rho & = & \bar{L}^{\phi} \hat{\rho}, \label{eq:r_rho} \\
\mathbf{v} & = & \bar{L}^{\delta} \hat{\mathbf{v}}, \label{eq:r_v} \\ 
d \hat{t} & = & \bar{L}^{\tau} d t, \label{eq:r_t}
\end{eqnarray}
in the density and momentum equations, Eqs~(\ref{eq:density},\ref{eq:f_momentum}), gives
\begin{align}
\begin{split} \label{eq:general_scaled_density}
\frac{\partial \hat{\rho}}{\partial \hat{t}} ={}& - \bar{L}^{\delta - \tau - 1} \nabla \cdot \left( \hat{\rho} \hat{\mathbf{v}} \right) - \bar{L}^{-\tau} \left(3 + \phi \right) H \hat{\rho}
\end{split} \\
\begin{split} \label{eq:general_scaled_momentum}
\frac{\partial \hat{\mathbf{v}}}{\partial \hat{t}} ={}& -\bar{L}^{\delta-\tau-1} \hat{\mathbf{v}}\cdot \nabla \hat{\mathbf{v}} - \bar{L}^{-\delta - \tau + \phi (\gamma -1) - 1} \frac{K}{\hat{\rho}} \nabla \hat{\rho}^{\gamma} \\
& - (1 + \delta) \bar{L}^{-\tau} H \hat{\mathbf{v}} + \bar{L}^{-\phi - \tau - 2} \hat{\mathbf{D}}.
\end{split}
\end{align}
The Hubble parameter $H = \dot{L}/L$, and the dissipation $\hat{\mathbf{D}}$ is the same as in Eq.~(\ref{eq:dissipation}), but with $\hat{\rho},\hat{\mathbf{v}}$ in place of $\rho,\mathbf{v}$.

By choosing $\phi = -3$ and $\delta = -1$, the forcing terms can be eliminated from the density and momentum equations. Then, choosing $\tau = -2$ removes the time-dependent coefficient from the divergence term in the density equation, and also removes it from the nonlinear term in the momentum equation. For these choices of $\phi,\delta,\tau$, the incompressible case of this transformation has been discussed by~\citet{cambon1992}. A different choice was made by~\citet{davidovits2016a,davidovits2016b}, for the convenience of simulations. Various similarity transformations (e.g. \citet{nishitani1985,nishitani1991,davis1977}) are related.

We will also take the polytropic index $\gamma = 5/3$. Then, the rescaled NS equations become,
\begin{align}
\begin{split} \label{eq:scaled_density}
\frac{\partial \hat{\rho}}{\partial \hat{t}} ={}& - \nabla \cdot \left( \hat{\rho} \hat{\mathbf{v}} \right)
\end{split} \\
\begin{split} \label{eq:scaled_momentum}
\frac{\partial \hat{\mathbf{v}}}{\partial \hat{t}} ={}& \hat{\mathbf{v}}\cdot \nabla \hat{\mathbf{v}} - \frac{K}{\hat{\rho}} \nabla \hat{\rho}^{5/3} + \bar{L}^{3} \hat{\mathbf{D}}.
\end{split}
\end{align}
Up to the $\bar{L}^3$ scaling on the dissipation term, these are the usual, unforced, NS equations for a gas with polytropic index $\gamma = 5/3$. Note that there is no separate energy equation because the system was closed with the assumption of a polytropic pressure, Eq.~(\ref{eq:polytropic}). An energy equation can be derived as usual from the momentum equation, Eq.~(\ref{eq:scaled_momentum}), but it does give ``new'' information, in the sense that the system is closed without it.

\section{Bound} \label{sec:bound}
Turbulence governed by the rescaled equations, Eqs.~(\ref{eq:scaled_density},\ref{eq:scaled_momentum}), will decay, as it is unforced in these variables. The usual compressible NS equations are recovered by setting $\bar{L} = 1$. For contraction $\bar{L}\!\left(\hat{t}\right) \leq 1$ is a strictly decreasing function of time (the equality holds at $\hat{t} = 0$). Since the viscous dissipation in Eq.~(\ref{eq:scaled_momentum}) is multiplied by $\bar{L}^3$, it has a smaller coefficient at all times after $\hat{t}=0$ in the compressing case than in the $\bar{L}=1$ usual case. Then it is reasonable to expect that the turbulent decay rate for the system Eqs.~(\ref{eq:scaled_density},\ref{eq:scaled_momentum}) will be slower than (or equal to) the decay for the usual compressible NS equations ($\bar{L}=1$). This is the key assumption on which the bound rests.

If this assumption is true, then the rms turbulent velocity for the system, Eqs.~(\ref{eq:scaled_density},\ref{eq:scaled_momentum}), will be at least as great as that given by the usual power-law decay for the system when $\bar{L}=1$,
\begin{equation}
\hat{v}_{rms} \geq \hat{v}_{rms,0} \left( 1 + \hat{t}/t_0 \right)^{-\alpha/2}. \label{eq:hat_bound}
\end{equation}
Here, $\alpha$ and $t_0$ are to be determined for turbulent decay in the non-compressing ($\bar{L}=1$) case. Then, arriving at the bound, Eq.~(\ref{eq:bound}) requires using Eqs.~(\ref{eq:r_v},\ref{eq:r_t}) to transform Eq.~(\ref{eq:hat_bound}) into the unscaled (non-hat) variables. One could instead write a comparable bound for the turbulent kinetic energy (TKE), $\langle \rho v^2 /2 \rangle$, under compression. We use the turbulent velocity in keeping with previous work \citep{robertson2012}. If a decay law of a different form than that given by Eq.~(\ref{eq:hat_bound}) is more appropriate, there will still be an equivalent bound, derived once again by transforming the decay law back into the unscaled variables.

Although we are not aware of work determining $t_0$ and $\alpha$ for the rms velocity decay of supersonic turbulence with $\gamma = 5/3$, we can estimate these values from closely related work. The bound will then be only a guide. \citet{maclow1999} found that for supersonic (initially mach 5) isothermal decaying turbulence, $t_0$ is the initial turnover time for the turbulence (at the driving scale). \citet{maclow1998} found that the TKE in supersonic (mach 5) turbulence with $\gamma = 7/5$ decays with power $\alpha \sim 1.2$. In the isothermal case ($\gamma = 1$), they found $\alpha \sim 1$, suggesting some slight dependence on $\gamma$, at least within this modest range. \citet{smith2000} found that for the decay of hypersonic (mach 50) isothermal turbulence, the decay power $\alpha \sim 1.5$, after an initial transient period. While these results suggest a single value of $\alpha$ will not suffice for all situations, we may expect that for $\gamma = 5/3$, $\alpha$ is roughly in the range $1 \sim 1.5$, depending on the initial mach number.

Note that these decay rates are for the TKE, not the rms velocity. Using them for the decay of the rms velocity discounts density-velocity correlations. \citet{maclow1999} found these correlations make for a $10\%-15\%$ difference between the TKE calculated from the rms velocity, $m v_{rms}^2/2$, and the TKE calculated directly, $\langle \hat{\rho} \hat{v}^2/2 \rangle$. Again, this result is for mach 5 turbulence, and may change with mach number.

\section{Discussion} \label{sec:discussion}

The bound, Eq.~(\ref{eq:bound}), can be used as a validation tool. For example, let us compare it with the compressing turbulence model and matching simulations of \citet{robertson2012}. That model is
\begin{equation}
\frac{d v_{rms}}{d \bar{L}} = - \left[1 + \eta \frac{v_{rms}}{H \bar{L} L_0} \right] \frac{v_{rms}}{\bar{L}}. \label{eq:robertson_model}
\end{equation}
This model for $v_{rms}$ includes two components: the forcing due to the contraction (the first term left of the equals sign in Eq.~(\ref{eq:momentum})), and the dissipation of $v_{rms}$ calculated from the equilibrium dissipation rate for forcing at a given scale, as found by \citet{maclow1999}. The forcing scale is taken to decrease in time as determined by the contraction, $\bar{L}$. \citet{robertson2012} found that $\eta = 1.2$ creates a good match between the model and their simulation results, which were carried out for isothermal turbulence. The model is nominally independent of $\gamma$, although to the extent the turbulent dissipation rate depends on $\gamma$, one may expect that $\eta$ could change.

Assuming equality in the bound, Eq.~(\ref{eq:bound}) and differentiating with respect to $\bar{L}$, one can write an expression for $v_{rms}$ that is comparable to Eq.~(\ref{eq:robertson_model}):
\begin{equation}
\frac{d v_{rms}}{d \bar{L}} = - \left[1 + \frac{\alpha}{2 t_0} \frac{\bar{L}^{2/\alpha - 2}}{H} \left(\frac{v_{rms}}{v_{rms,0}} \right)^{2/\alpha} \right] \frac{v_{rms}}{\bar{L}}. \label{eq:equality}
\end{equation}
For $\alpha = 2$, $\eta = 1$, and taking the initial turnover time as $t_0 = L_0/v_{rms,0}$, the two expressions, Eq.~(\ref{eq:robertson_model}) and Eq.~(\ref{eq:equality}), are equal. The bound and the model of \citet{robertson2012} are calculated for different values of $\gamma$, complicating a direct comparison. To compare, consider the $\gamma = 5/3$ case, for which the bound is calculated. First, assume that the model, Eq.~(\ref{eq:robertson_model}), is still valid for $\gamma \neq 1$, with a possibly different value of $\eta$ (as suggested by \citet{robertson2012}). Then, the comparison hinges on the value of the decay power $\alpha$ of $v_{rms}$ for supersonic turbulence when $\gamma = 5/3$.

To see this, consider the asymptotic scaling of $v_{rms}$ with $\bar{L} \rightarrow 0$, for the case of a gravitational-like contraction with $H = -H_0 \bar{L}^{-3/2}$. The prediction of the model, Eq.~(\ref{eq:robertson_model}), is
\begin{equation}
v_{rms}/v_{rms,0} \rightarrow (H_0 L_0/2 v_{rms,0} \eta) \bar{L}^{-1/2}. \label{eq:model_limit}
\end{equation}
Using Eq.~(\ref{eq:equality}) with $t_0 = L_0/v_{rms,0}$, the same contraction gives for $\bar{L} \rightarrow 0$ that
\begin{equation}
v_{rms}/v_{rms,0} \rightarrow (H_0 L_0/2 v_{rms,0})^{\alpha/2} \bar{L}^{\alpha/4 - 1}. \label{eq:bound_limit}
\end{equation}
Now, note that, unless $\alpha \geq 2$, the model Eq.~(\ref{eq:robertson_model}) will cross below the bound as $\bar{L} \rightarrow 0$ for \emph{any} $\eta$. Apparently, assuming the bound holds, either $\alpha \geq 2$ for the decay of $v_{rms}$ when $\gamma = 5/3$, or the model, Eq.~(\ref{eq:robertson_model}) will be too dissipative when applied to the $\gamma = 5/3$ case.

We are not aware of available decay rates for $v_{rms}$ in turbulence with $\gamma = 5/3$ and initial mach numbers $M \sim 6$. However, available decay rates in the literature, for various values of $\gamma$ and initial mach number suggest $\alpha \geq 2$ would be an outlier. In the low-mach case, decay rates this large are associated with bounded turbulence \citep{skrbek2000,thornber2016}. \citet{maclow1998} found for initially mach 5 turbulence a TKE decay rate $\alpha \sim 1$ for $\gamma = 1$ and $\alpha \sim 1.2$ for $\gamma = 7/5$. For initially mach-20 turbulence with a complicated equation of state, \citet{pavlovski2002,pavlovski2006} find for the TKE that $\alpha \sim 1.34$. These decay rates are calculated for the TKE. As noted in Sec.~\ref{sec:bound}, using them for $v_{rms}$ neglects density-velocity correlations. While the size of these correlations will likely change when $\gamma$ changes, they have a relatively small impact when $\gamma = 1$. The decay rate of $v_{rms}$ has been found directly, using a number of different simulation algorithms, by \citet{kitsionas2009}. They found for isothermal, mach 4 turbulence, $\alpha \sim 1$, which is very similar to the decay rate inferred from the TKE. Given the available results, it seems unlikely the bound, Eq.~(\ref{eq:bound}) and the model, Eq.~(\ref{eq:robertson_model}) will be consistent for the $\gamma = 5/3$ case, with the model being too dissipative. As the methods used to arrive at both the model and the bound appear reasonable, reconciling this difference requires more detailed consideration.

Assuming equality in the bound, Eq.~(\ref{eq:bound}), is equivalent to asserting that the time-dependent pre-factor ($\bar{L}^3$) of the viscous dissipation, $\hat{\mathbf{D}}$, in Eq.~(\ref{eq:scaled_momentum}) does not decrease the dissipation rate of the turbulence, despite the fact that the coefficient decreases in time. That is, that the dissipation rate (and therefore energy behavior) of the turbulence is independent of time dependence in the viscous coefficient. For various subsonic compressing turbulence studies, this has not been found to be the case \citep{coleman1991,cambon1992,davidovits2016b}. Since dissipation in decaying supersonic (isothermal) turbulence is primarily in shocks \citep{smith2000}, it is conceivable that the situation changes between subsonic and supersonic turbulence. Perhaps more importantly, in the previously studied subsonic cases, the viscous coefficient was generally increasing in time, rather than decreasing as in the present situation.

If the shrinking-in-time dissipation coefficient did have no impact on the dissipation rate, then Eq.~(\ref{eq:bound}), with equality assumed, would be a model for $v_{rms}$, rather than a bound. Furthermore, in this case, for a given initial condition, a single simulation of Eqs.~(\ref{eq:scaled_density},\ref{eq:scaled_momentum}) would be sufficient for \emph{all} compression histories $\bar{L}\!\left(t\right)$ (or Hubble parameters, $H\!\left(t\right)$, alternatively). This is because the Eqs.~(\ref{eq:scaled_density},\ref{eq:scaled_momentum}) would no longer have any dependence on $\bar{L}$. 

If the shrinking dissipation coefficient counter-intuitively led to \emph{more} dissipation than in the case where the coefficient is constant, the lower bound would be invalid. If this effect were consistent, it would instead represent an upper bound (with $\geq$ in Eq.~(\ref{eq:bound}) switching to $\leq$). 

The bound depends to some degree on the choice of physical model for the dissipation process. If the relevant dissipation process is not captured by the NS viscous dissipation, Eq.~(\ref{eq:dissipation}), the bound may change. This is because the time-dependent coefficient of the dissipation in the rescaled momentum equation, Eq.~(\ref{eq:scaled_momentum}), results from transformation and rescaling of the dissipation. For a different dissipation form, one could imagine that the coefficient after rescaling is different from the $\bar{L}^3$ coefficient found here. This could alter the bound, particularly if the coefficient were no longer shrinking in time. Additionally, the form of the NS dissipation, $\mathbf{D}$, does not change under transformation to the moving frame and rescaled variables, allowing the analogy between the compressing case and the uncompressing case.

This will not necessarily be true for all imaginable dissipation forms. For example, if the physically correct dissipation for the fluid equations took the form of the artificial viscosity commonly used for shock-capturing (see e.g. \citet{vonneumann1950,stone1992}), the bound would need to be reconsidered. Note that, for numerical simulations in the moving frame (solving Eqs.~(\ref{eq:density},\ref{eq:momentum})), the form of the dissipation may need to be considered explicitly, as done here, so that its transformation can be accounted for.

We now turn to the dependence of the bound on the adiabatic index. When $\gamma \neq 5/3$, the scaled momentum equation, Eq.~(\ref{eq:scaled_momentum}), will pick up additional time dependence, as a coefficient for the pressure gradient term. This worsens the analogy between the scaled momentum equation and regular NS, but need not necessarily dramatically alter the bound. The impact on the bound depends on the effect of the pressure term (through the pressure-dilatation) on the energy dissipation in supersonic turbulence. As an example, consider the isothermal scenario ($\gamma =1$). In this case, the scaled momentum equation becomes
\begin{equation}
\frac{\partial \hat{\mathbf{v}}}{\partial \hat{t}} = \hat{\mathbf{v}}\cdot \nabla \hat{\mathbf{v}} - \bar{L}^2 \frac{K}{\hat{\rho}} \nabla \hat{\rho} + \bar{L}^{3} \hat{\mathbf{D}}. \label{eq:scaled_isothermal}
\end{equation}
The decay rate of compressible turbulence is the result of the net effect of the viscous dissipation, $\propto \hat{\mathbf{v}}\cdot\hat{\mathbf{D}}$, and the pressure-dilatation, which comes from the dot product of the pressure gradient term with $\hat{\mathbf{v}}$. In the high-Reynolds-number limit, it can be shown that the mean pressure-dilatation acts primarily on the largest scales, with its impact on small scales averaging out \citep{aluie2011,aluie2013,aluie2012}. To the extent that the pressure-dilatation enhances the decay rate, the bound should be insensitive to the $\bar{L}^2$ scaling. This is because the bound comes about by considering $\bar{L} = 1$ to be a more dissipative case than when $\bar{L}$ shrinks in time, which would remain the case. There is some evidence the pressure-dilatation does in fact increase the dissipation in decaying turbulence, at least in the subsonic case \citep{sarkar1992,samtaney2001}. Even without this, the bound will be approximately preserved so long as the net effect on the decay of the pressure-dilatation term with the $\bar{L}^2$ coefficient is small compared to that of the viscous dissipation term with the $\bar{L}^3$ coefficient. For the isothermal case the pressure term scales as the sound speed squared, $C_s^2$, which becomes small in the high-mach limit. However, for very large compressions (reaching very small $\bar{L}$), the weaker decrease on the pressure term may relatively enhance its contribution even if it would normally be small.

Overall, even for $\gamma = 5/3$, the bound can only be universal to the extent that the decay of supersonic turbulence is \citep{federrath2013}. To the extent the mix of compressible and solenoidal modes in the initial condition affects the decay rate, this must be accounted for in the value of $\alpha$. Similarly with the impact of changing $\gamma$ and changing initial mach number.

As noted in Sec.~\ref{sec:model}, the present treatment considers contractions where the time dependence of $\bar{L}$ is determined by the gravitational attraction of the mean density. Strictly speaking, if $\bar{L}$ is taken to have a different form, one must consider the effect of an acceleration term $\ddot{\bar{L}} \mathbf{x}$ in the momentum equation, Eq.~(\ref{eq:momentum}). This may or may not have a significant impact on the bound. As also noted in Sec.~\ref{sec:model}, gravitational effects from the density fluctuations have been neglected. In many astrophysical problems of interest, there is forcing besides for the contraction which acts on the turbulence, which is neglected here. 

These various assumptions and restrictions, if limiting to the generality of the bound, should be replicable for simulations. Then the bound provides a relatively simple, high-level check on the simulations, particularly on the degree of dissipation. The initial application of the bound in this manner suggests a commonly used model (and matching simulations) may be too dissipative. Note that, even if a simulation or model is too dissipative, it may still be useful, depending on the physics under consideration. 

The implications of the rescaled equations, Eqs.~(\ref{eq:scaled_density},\ref{eq:scaled_momentum}), apart from the bound, deserve mention. These equations are reached because forcing of the type generated by contraction can be scaled out of the NS equations. The only difference between the rescaled equations and compressible NS equations is in the dissipation term. However, many turbulent quantities, for example, inertial range properties, are not influenced by the dissipation properties of the turbulence. Therefore, we may expect that already known results for decaying supersonic turbulence can be translated by undoing the scaling, and applied to turbulence undergoing compression. This task is made simpler by the fact that the rescaling, Eqs.~(\ref{eq:r_rho},\ref{eq:r_v},\ref{eq:r_t}) is purely time dependent, so that, for example, spatial correlation functions are translatable.

In conclusion, we have suggested a lower bound on the increase in turbulent velocity associated with compression of turbulence. This lower bound follows directly once one assumes that a decreasing-in-time coefficient of viscosity in the NS equations does not increase the rate of dissipation for turbulence. This assumption, while physically reasonable, should be verified or disproved, since the bound represents a useful means of checking models or simulations of compressing turbulence, and an initial application of the bound in this capacity indicates some previous work may be too dissipative.

\begin{acknowledgments}
This work was supported by NNSA 67350-9960 (Prime $\#$ DOE DE-NA0001836), NNSA SSAA Grant DE-NA0002948 and by NSF Contract No. PHY-1506122.
\end{acknowledgments}

\bibliographystyle{apj}

\end{document}